\documentclass[twocolumns]{emulateapj}
\usepackage{natbib}
\usepackage{amssymb}
\usepackage{amsmath}
\def\be{\begin{eqnarray}}
\def\ee{\end{eqnarray}}
\usepackage{comment}
\usepackage[dvipsnames]{xcolor}
\usepackage{graphicx,subfigure}
\usepackage{enumerate}
\usepackage{rotating}
\usepackage{longtable}
\usepackage{booktabs}
\usepackage{float}

\usepackage{hyperref}
\hypersetup{colorlinks,urlcolor={blue},linkcolor={blue},citecolor={blue}}

\shorttitle{Radio follow-up of GW170817}

\shortauthors{}

\begin{document}

\title{GW170817 4.5 years after merger: Dynamical ejecta afterglow constraints}
\author{Arvind~Balasubramanian\altaffilmark{1}}
\author{Alessandra~Corsi\altaffilmark{1}}
\author{Kunal~P.~Mooley\altaffilmark{2}}
\author{Kenta~Hotokezaka\altaffilmark{3}}
\author{David~L.~Kaplan\altaffilmark{4}}
\author{Dale~A.~Frail\altaffilmark{5}}
\author{Gregg~Hallinan\altaffilmark{2}}
\author{Davide~Lazzati\altaffilmark{6}}
\author{Eric~J.~Murphy\altaffilmark{7}}
\altaffiltext{1}{Department of Physics and Astronomy, Texas Tech University, Box 1051, Lubbock, TX 79409-1051, USA; e-mail: arvind.balasubramanian@ttu.edu}
\altaffiltext{2}{Caltech, 1200 E. California Blvd. MC 249-17, Pasadena, CA 91125, USA}
\altaffiltext{3}{Research Center for the Early Universe, Graduate School of Science, University of Tokyo, Bunkyo-ku, Tokyo 113-0033, Japan}
\altaffiltext{4}{Center for Gravitation, Cosmology, and Astrophysics,
Dept. of Physics, University of Wisconsin-Milwaukee, P.O. Box 413, Milwaukee, WI 53201, USA}
\altaffiltext{5}{National Radio Astronomy Observatory, Socorro, NM 87801, USA}
\altaffiltext{6}{Department of Physics, Oregon State University, 301 Weniger Hall, Corvallis, OR 97331, USA}
\altaffiltext{7}{National Radio Astronomy Observatory, Charlottesville, VA 22903, USA}

\begin{abstract}
GW170817 is the first binary neutron star (NS) merger detected in gravitational waves (GWs) and photons, and so far remains the only GW event of its class with a definitive electromagnetic (EM) counterpart. Radio emission from the structured jet associated with GW170817 has faded below the sensitivity achievable via deep radio observations with the most sensitive radio arrays currently in operation. Hence, we now have the opportunity to probe the radio re-brightening that some models predict, should emerge at late times from the interaction of the dynamically-stripped merger ejecta with the interstellar medium. Here we present the latest results from our deep radio observations of the GW170817 field with the Karl G. Jansky Very Large Array (VLA), 4.5 years after the merger. Our new data at 3\,GHz do not show any compelling evidence for emission in excess to the tail of the jet afterglow ($<3.3\,\mu$Jy), confirming our previous results. We thus set new constraints on the dynamical ejecta afterglow models. These constraints favor single-speed ejecta with energy $\lesssim 10^{50}$\,erg (for an ejecta speed of $\beta_0=0.5$), or steeper energy-speed distributions of the kilonova ejecta. Our results also suggest larger values of the cold, non-rotating maximum NS mass in equal mass scenarios. However, without a detection of the dynamical ejecta afterglow, obtaining precise constraints on the NS equation of state remains challenging. 
\end{abstract}


\section{Introduction}\label{sec:intro}
GW170817 remains the first and only example yet of the merger of two neutron stars (NSs), observed by LIGO and VIRGO  \citep{2017PhRvL.119p1101A}, whose discovery in gravitational waves (GWs) was followed by the identification of a definitive electromagnetic (EM) counterpart from radio to $\gamma$-ray frequencies. 

The treasure trove of information that this event has provided to the astronomy community cannot be understated. We refer the reader to the many papers written on this event for a comprehensive description of all of the observations that enabled the identification of a coincident gamma-ray burst \citep[GRB; e.g.][and references therein]{2017ApJ...848L..13A}; a host galaxy at 40\,Mpc and a UV/optical/IR kilonova   \citep[e.g.,][]{2017Natur.551...64A,2017ApJ...848L..19C,2017Sci...358.1556C,2017ApJ...848L..17C,2017Sci...358.1570D,2017Sci...358.1559K,2017Natur.551...80K,2017Sci...358.1583K,2017Natur.551...67P,2017Sci...358.1574S,2017Natur.551...75S,2017ApJ...848L..27T,2017ApJ...848L..24V,2017arXiv171005931M}; a delayed non-thermal afterglow observed from radio to X-rays  \citep[e.g.,][]{2017Natur.551...71T,2017ApJ...848L..25H,2017ApJ...848L..20M,2017Sci...358.1579H}. 

Extensive observations of the quasi-thermal kilonova and of the non-thermal afterglow associated with GW170817 have painted a detailed picture of the ejecta that resulted from the merger of the two NSs in the compact binary progenitor of GW170817. While the kilonova was powered by quasi-isotropic and relatively slow neutron-rich debris originating from a combination of dynamical ejecta and disk winds \citep[e.g.][]{2017arXiv171005931M}, the non-thermal radio afterglow probed the existence of an off-axis jet that successfully burrowed through the neutron-rich debris. Radio observations, in particular, were instrumental in narrowing down the morphology of relativistic ejecta to a structured jet (a.k.a. jet$+$cocoon), and in providing crucial insights into the geometry of the merger itself, and the density of the interstellar medium (ISM) through which the jet was launched \citep{2018Natur.561..355M,2018PhRvL.120x1103L,2020ApJ...901L..26R,2019Sci...363..968G,2018ApJ...861L..10C,2018ApJ...858L..15D,2017ApJ...848L..21A,2018ApJ...856L..18M,2018Natur.554..207M,2018ApJ...868L..11M,2019ApJ...886L..17H}. 

Well before the discovery of GW170817, models had been proposed predicting that, regardless of whether a jet is successfully launched in a binary NS merger, the interaction of the kilonova ejecta with the ISM can produce non-thermal emission  in the radio a few years after merger \citep[e.g.,][]{2011Natur.478...82N,2013MNRAS.430.2121P,2015MNRAS.450.1430H}, motivating several related observational efforts in cosmological short GRBs \citep[e.g.][]{2014MNRAS.437.1821M,2016ApJ...831..141F,2016ApJ...819L..22H}. With GW170817, these late-time re-brightening models have spurred new interest in the community \citep{2019MNRAS.487.3914K,2018ApJ...867...95H,2019MNRAS.485.4150B,2020MNRAS.495.4981M,2020ApJ...890..102L}, especially given their potential to probe the nature of the merger remnant in relation to the Equation of State (EoS) of nuclear matter \citep[see e.g.][and references therein]{Nedora2021}. Thus, additional observational campaigns have been carried out in search for late-time radio afterglows in both GW170817 \citep{2021ApJ...914L..20B,2022ApJ...927L..17H,2022MNRAS.510.1902T} and other short GRBs \citep[e.g.,][]{2019ApJ...887..206K,2021MNRAS.505L..41B,2021ApJ...908...63G,2021MNRAS.500.1708R}, albeit without any definitive detections so far. 

Motivated by the above considerations, in \citet{2021ApJ...914L..20B} we presented the deepest radio observations of the GW170817 field at 3.5\,years after merger, and found no evidence for a late-time radio re-brightening. This result helped constrain the energy-speed distribution of the kilonova ejecta \citep{2021ApJ...914L..20B}, and provided hints on the NS EoS \citep{Nedora2021}. On the other hand, late-time X-ray observations of the GW170817 field around the same epoch had left open the possibility of late-time emission in excess to that expected from the tail of the GW170817 jet afterglow \citep{2022ApJ...927L..17H,2022MNRAS.510.1902T}. However, continued follow-up in the X-rays at 4.3-4.8 years since merger did not confirm the presence of an X-ray excess at these later times \citep{2021GCN.31231....1H,Oconnor_GCN}.  

Here, we present new deep observations of the GW170817 field carried carried out with the Karl G. Jansky Very large Array (VLA) at 3\,GHz and at the epoch of about 4.5\,years since merger. These observations improve substantially on the sensitivity reached by recently reported radio observations of the same field \citep{Ricci_GCN}. 
Our paper is organized as follows. We report our new observations in $\S$\ref{sec:obs}; in $\S$\ref{sec:disc} we discuss our results within the kilonova ejecta afterglow model; finally, in $\S$\ref{sec:conc} we summarize and conclude.  


\setlength\LTcapwidth{\linewidth}
\begin{longtable*}{cccccccc}
\caption{VLA late-time observations of the GW170817 field. See text for details on RMS measurements.
\label{tab:rad_obs}}\\
\toprule
\toprule
Date  & $\nu$  & VLA & Time on-source & RMS & VLA   & PI & Nominal synth. beam\\
(UT) & (GHz) & config. & (hr) & ($\mu$Jy) & program & &(\arcsec)\\
\midrule
2021 Dec 06 & 3.0 & B & 2\,h24\,m54\,s & 4.9 (3.6) & 21B-057 &  Balasubramanian & 2.1\\
2021 Dec 20 & 3.0 & B & 2\,h25\,m57\,s & 4.4 (3.9) & 21B-057 &  Balasubramanian & 2.1\\
2021 Dec 28 & 3.0 & B & 2\,h25\,m57\,s & 4.7 (3.9) & 21B-057 &  Balasubramanian & 2.1\\
2022 Jan 05 & 3.0 & B & 2\,h25\,m57\,s & 4.9 (4.3) & 21B-057 &  Balasubramanian & 2.1\\
2022 Mar 05 & 2.9 & A & 2\,h28\,m00\,s & 4.4 (3.9) & 22A-168 & Balasubramanian & 0.65\\
2022 Mar 10 & 3.0 & A & 2\,h28\,m08\,s & 4.4 (3.9) & 22A-168 & Balasubramanian & 0.65\\
2022 Mar 14 & 3.0 & A & 2\,h28\,m02\,s & 4.0 (3.9) & 22A-168 & Balasubramanian & 0.65\\
2022 Mar 17 & 3.0 & A & 2\,h28\,m08\,s & 4.0 (3.9) & 22A-168 & Balasubramanian & 0.65\\
2022 Mar 22 & 3.0 & A & 2\,h28\,m10\,s & 4.1 (4.0) & 22A-168 & Balasubramanian & 0.65\\
2022 Mar 23 & 3.0 & A & 2\,h17\,m34\,s & 4.4 (4.1) & 22A-168 & Balasubramanian & 0.65\\
2022 Mar 28 & 3.0 & A & 2\,h38\,m42\,s & 3.7 (3.4) & 22A-168 & Balasubramanian & 0.65\\
2022 Mar 29 & 3.0 & A & 2\,h28\,m00\,s & 3.9 (3.7) & 22A-168 & Balasubramanian & 0.65\\
\bottomrule
\end{longtable*}

\setlength\LTcapwidth{\linewidth}
\begin{longtable*}{ccccccc}
\caption{Results for the co-added late-time radio observations of GW170817. See text for discussion.
\label{tab:obs}}\\
\toprule
\toprule
Date  & Epoch & $\nu$ & $F_{\nu}$ & $\sigma_{\nu}$ & Instrument & Reference\\
(UT) & (yr) & (Hz) & ($\mu$Jy) & ($\mu$Jy)& &\\
\midrule
2021 Dec 06 - 2022 Jan 05 & 4.3 & $2.8\times10^9$ & $<6.6$ & 2.2  & VLA B & This work\\
2022 Mar 05 - 2022 Mar 29 & 4.6 & $3.0\times10^9$ & $<4.5$ & 1.5   & VLA A & This work\\
2021 Dec 06 - 2022 Mar 29 & 4.5 & $3.0\times10^9$ & $<3.3$ & 1.1  & VLA A\&B & This work\\
2021 Dec 07 - 2022 May 18 & 4.5 & $2.41\times10^{17}$ & $5.18\times10^{-5}$ & $3.44\times10^{-5}$ & Chandra& \cite{Oconnor_GCN}\\
\bottomrule
\end{longtable*}
\section{Observations and data reduction}\label{sec:obs}

\begin{figure}
    \centering
    \includegraphics[width=\columnwidth]{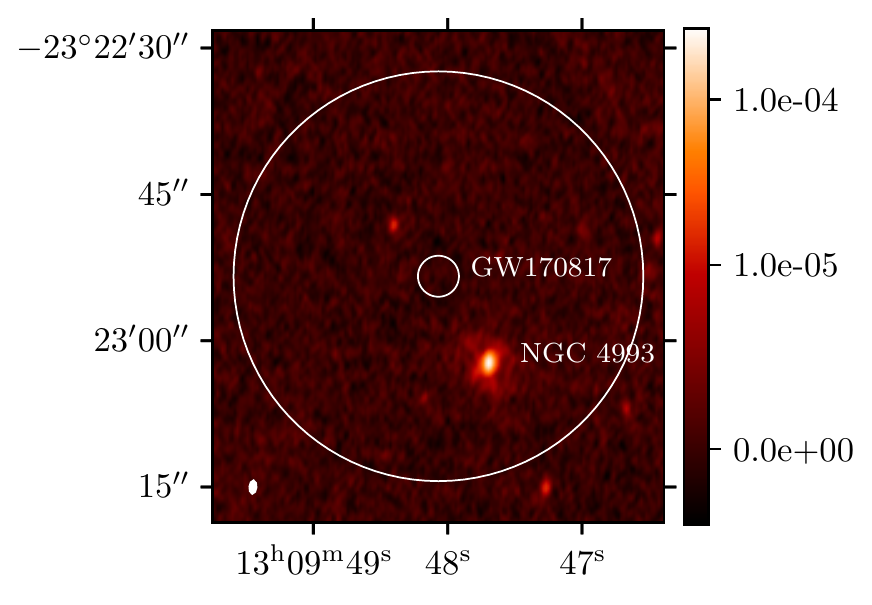}
    \caption{Image of the GW170817 field at $\approx 4.5$\,yrs since merger, as derived from our deepest co-added dataset  (see Table \ref{tab:obs}). The small circle has a radius of $2.1\,\arcsec$ and is centered on the position of GW170817. The larger circle has a radius of $21\,\arcsec$, equal to the radius of the circular region used to calculate our RMS sensitivity in the residual image of the field. The host galaxy of GW170817 is enclosed in this larger circular region. Several sources unrelated to GW170817 are also visible. The synthesized beam ellipse is shown in the bottom left. The color bar gives the flux density in Jy. 
    }
    \label{fig:GW170817_deep_image}
\end{figure}

We carried out radio continuum observations of the GW170817 field with the VLA. Our observations were executed with the standard VLA S band setup, with a nominal central frequency of 3\,GHz, and split in 12 epochs (each providing approximately 2.5\,hours on source) between December 2021 and March 2022. The first four epochs were observed with the VLA in its B configuration, while the subsequent eight epochs were carried out with the array in its most extended A configuration. These observations are listed in Table \ref{tab:rad_obs}. After calibration was performed with the automated VLA calibration pipeline, we manually inspected the data and performed further flagging for radio frequency interference (RFI) as needed. We then imaged the data using the \texttt{CASA} \citep{2007ASPC..376..127M} task \texttt{tclean} with one Taylor term (\texttt{nterms=1}) and robust weighting \citep[\texttt{robust=0.5}; see also][]{2021ApJ...914L..20B}, and derived the sensitivity RMS measurements running \texttt{imstat} on the residual images within a circular region of radius equal to 10 nominal synthesized beams\footnote{As recommended by \citet{hancock2012} and \citet{mooley2013}.} around the position of GW170817  \citep[$\alpha=13{\rm h}09{\rm m}48.069{\rm s}$, $\delta=-23{\rm d}22{\rm m}53.39{\rm s}$, J2000;][]{2018Natur.561..355M}. Because this region may include residuals associated with the host galaxy light (see Figure \ref{fig:GW170817_deep_image}), we also list in parenthesis in Table \ref{tab:rad_obs} the RMS values we obtain using a circular region of the same size in a source-free portion of the image. We find no significant ($>3\times$RMS) excess in a region of one synthesized beam around the position of GW170817 in any of the individual epochs. 

Next, we co-add the four B configuration observations, and the eight A configuration observations separately; finally, we co-add the full multiple configuration data set (all in the visibility domain) for a total of 12 observations. The imaging for these co-added datasets was performed similar to what is described above, with the \texttt{CASA} task \texttt{tclean} but using \texttt{nterms=2} to better clean the emission from bright radio sources in the field. To estimate the RMS sensitivity for the co-added observations in the A and B configurations, we conservatively use a circular region of radius 10 times the nominal synthesized beam width of the B configuration, centered on the location of GW170817 in the residual images (Figure 1 and Table \ref{tab:obs}).  We note that the RMS values estimated this way differ by less than 10\% from RMS values calculated in source free regions of the cleaned image. In our deepest co-added image we reach an RMS sensitivity of $1.1\,\mu$Jy at 3.0\,GHz. No emission in excess to $3\times$ the co-added image RMS is found in a circular region of radius 2.1\,arcsec (FWHM of the nominal VLA synthesized beam in B configuration at 3\,GHz) around the location of GW170817. Specifically, at the location of GW170817 we measure a 3\,GHz flux of $2.1\pm1.1\,\mu$Jy. Therefore, we constrain the radio emission from GW170817 to $<3.3\,\mu$Jy at 4.5\,years since merger (see Figure \ref{fig:GW170817_lc}).  

\section{Discussion}\label{sec:disc}
\begin{figure*}
    \centering
    \includegraphics{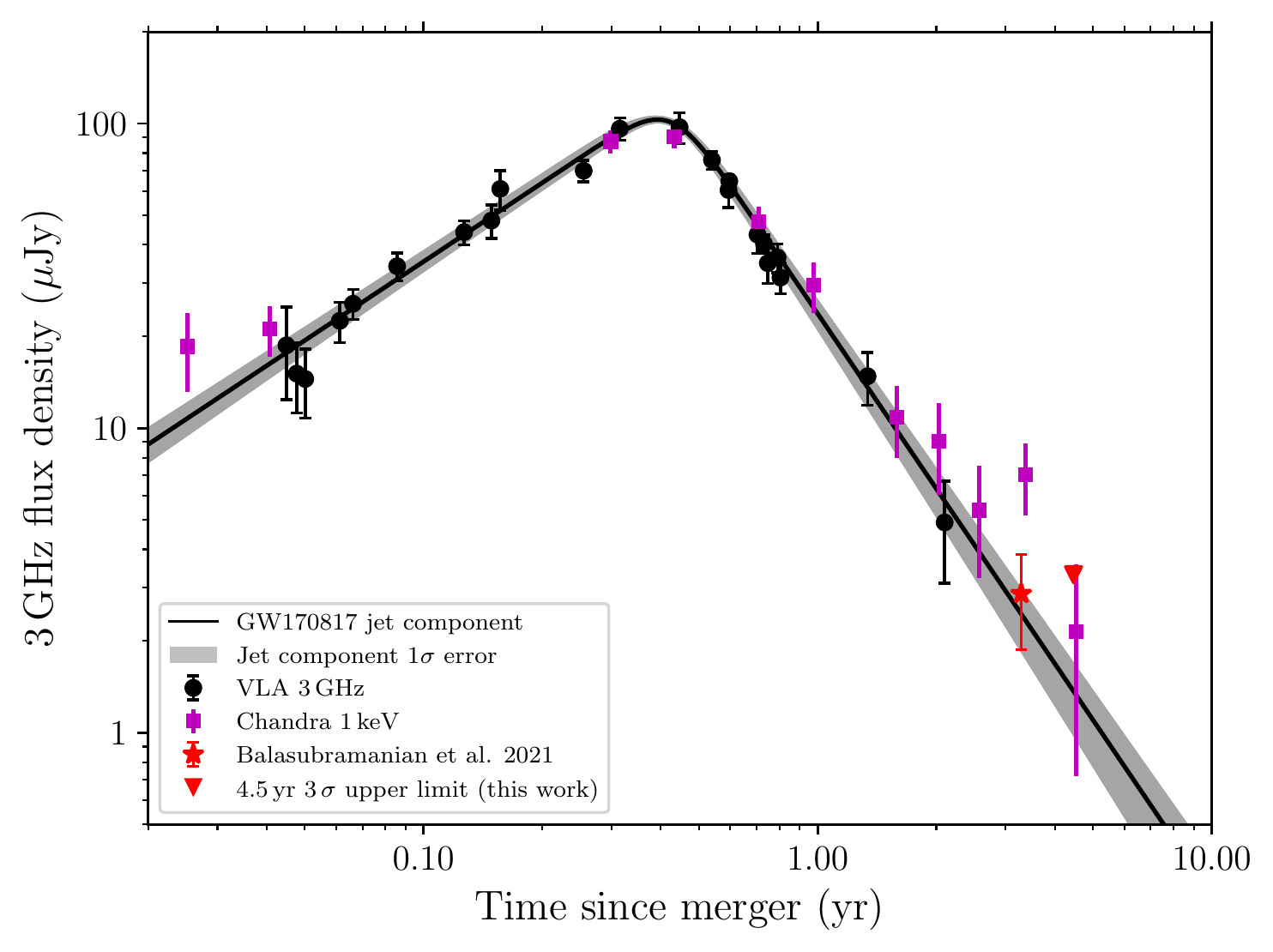}
    \vspace{-0.2cm}
    \caption{3\,GHz radio light curve of GW170817 with the best fit structured jet model from \cite{2021ApJ...922..154M}. Radio data are shown as black data points. The \textit{Chandra} 1\,keV data  scaled to 3\,GHz with a power-law index of $\beta=-0.584$  \citep[including the latest measurement by][]{Oconnor_GCN} are shown as purple squares \citep[see e.g.][and references therein]{2017ApJ...848L..25H,2017ApJ...848L..20M,2017Natur.551...71T,2022ApJ...927L..17H,2022MNRAS.510.1902T}. We  extrapolate all X-ray data to 3\,GHz using $\beta=-0.584$ because there is only marginal evidence for a potential spectral flattening around 3.5 yrs since merger \citep{2021ApJ...914L..20B,2022ApJ...927L..17H} which, however, has not been confirmed in later observations by \citet{Oconnor_GCN}. Our previous observation  3.5\,years since merger is marked with a red star \citep{2021ApJ...914L..20B}. The 3\,$\sigma$ upper limit from this work is shown as red downward pointing triangle.}
    \label{fig:GW170817_lc}
\end{figure*}

In Figure \ref{fig:GW170817_lc}, we show the 3\,GHz light curve of GW170817 (see the panchromatic afterglow data webpage\footnote{\url{http://www.tauceti.caltech.edu/kunal/gw170817/gw170817_afterglow_data_full.txt} and \url{https://github.com/kmooley/GW170817/}} for a compilation of the full dataset). The black data points are the previous radio observations \citep[][]{2017Sci...358.1579H,2018Natur.554..207M,2018ApJ...868L..11M,2021ApJ...922..154M} that follow the jet$+$cocoon afterglow model (black line with gray 1$\sigma$ error region). The red star shows our previous radio detection at $\sim3.5$\,years since merger \citep{2021ApJ...914L..20B}. The radio upper limit from this work is shown with a downward pointing red triangle.  As evident from this Figure, we do not find any significant evidence for emission in excess to the expectations from a decaying jet$+$cocoon afterglow model, confirming our previous results \citep{2021ApJ...914L..20B}.

For comparison, in Figure \ref{fig:GW170817_lc} we also show the X-ray flux measurements derived from \textit{Chandra} observations of the GW170817 field are shown as purple squares \citep[see e.g.][and references therein]{2017ApJ...848L..25H,2017ApJ...848L..20M,2017Natur.551...71T,2022ApJ...927L..17H,2022MNRAS.510.1902T} extrapolated to the radio band  using a radio-to-X-ray spectral index of $\beta=-0.584$ \citep[see][]{2021ApJ...922..154M}. Recently, \cite{Oconnor_GCN} also reported a measurement of $\sim0.6\times10^{15}\,\rm{erg}\,\rm{cm}^{-2}\rm{s}^{-1}$ for the $0.3-10$\,keV flux of GW170817 at $\approx 4.8$\,years after the merger (assuming a spectral index of $-0.585$), using observations carried out with the \textit{Chandra} observatory \citep[][]{Oconnor_GCN,2021GCN.31231....1H}. We convert this flux into a flux density at 1\,keV (see Table \ref{tab:obs}) and, by combining it with the radio upper-limit presented here, we derive a  radio-to-X-ray spectral index of $\beta \gtrsim -0.608$. This is compatible with the best fit value obtained via previous  observations of the structured jet afterglow \citep[$\beta=-0.584\pm0.002$;][]{2021ApJ...922..154M}, and with the results of our analysis at 3.5\,years after merger  \citep[$\beta=-0.535\pm0.024$ ;][]{2021ApJ...914L..20B}.

Hereafter we discuss the implications of our latest radio observations in the context of the kilonova ejecta model, following the formulation of \citet{2019MNRAS.487.3914K}. In this model, the kilonova blast wave drives a shock through the ISM, resulting in synchrotron emission. Electrons are accelerated to a power-law distribution of Lorentz gamma factors $\gamma_{e} > \gamma_{e,m}$, with power-law index $p$. The energy in the kilonova spherical blast wave is distributed as $E(>\beta \gamma)\propto(\beta\gamma)^{-\alpha}$ (with $\gamma$ being the Lorentz factor, $\beta$ being the speed in units of speed of light of the shocked fluid and $\alpha$ being the power law index of the energy speed distribution) and normalized to  the  total energy  $E$  at  some  minimum  velocity $\beta_0$ such that $E(>\beta_0\gamma_0)=E$. It is reasonable to assume that radio (GHz) observations are in between the minimum frequency, $\nu_{m}$ \citep[corresponding to $\gamma_{m}$, see][] {2011Natur.478...82N}, and the cooling frequency, $\nu_{c}$.  In this case, the kilonova peak flux density reads \citep{2011Natur.478...82N}:

\begin{equation}\label{eq:peak_flux}
    F_{\rm \nu, pk}\approx (1.5\,{\rm mJy})\,\epsilon_{\rm e,-1}^{p-1}\,\epsilon_{\rm B,-3}^{\frac{p+1}{4}}\,n_{-2}^{\frac{p+1}{4}}\,\beta_{0}^{\frac{5p-7}{2}}\,E_{51}\,\nu_{9.5}^{\frac{1-p}{2}}\,d_{26}^{-2},
\end{equation}
where $Q_{x}=Q/10^{x}$ is followed for all quantities ($Q$, all expressed in cgs units); $\epsilon_{\rm B}$ and $\epsilon_{\rm e}$ are the fractions of the total energy in the magnetic field and electrons respectively; $n$, the number density of the medium; $d$ is the distance to the source; the normalization  constant is calculated for $p=2.1$. The time at which the kilonova afterglow emission peaks can be calculated as \citep{2019MNRAS.487.3914K}:
\begin{equation}
    t_{\rm dec}=t_{\rm pk}\approx (3.3 {\rm yr})\left(\frac{E_{\rm  51}}{n_{-2}}\right)^{\frac{1}{3}}\, \beta_{0}^{-\frac{2}{3}}\left(\frac{2+\alpha}{\beta_0(5+\alpha)}-1\right).
\end{equation}
where $\alpha$ is the power law index of the energy speed distribution discussed earlier. The blast wave can be approximated to be mildly relativistic before this peak, and therefore the rising part of the kilonova ejecta light curve can be easily modeled as \citep[see][and references therein]{2019MNRAS.487.3914K}:
\begin{equation}\label{eq:kn_rise}
    F_{\rm\nu, KN}\,(t)=F_{\rm \nu,pk}\left(\frac{t}{t_{\rm p}}\right)^s,
\end{equation}
where:
\begin{equation}
    s=\frac{3\alpha-6(p-1)}{8+\alpha}.
\end{equation}

For $\alpha=\infty$, Equations (1)-(3) reduce to the case of a spherical outflow of total energy $E$ with uniform velocity $\beta_0$  \citep{2011Natur.478...82N}. In this case, our flux upper-limit at 4.5\,yr constrains the energy $E$ and speed $\beta_0$ for a given choice of the density and micro-physical parameters. Indeed, setting these parameters as in \citet{2021ApJ...922..154M}, an energy of $E\approx 10^{50}\,$erg and speed of $\beta_{0}\approx 0.5$ would produce a radio peak flux comparable to our $3\sigma$ upper-limit at 4.5\,yrs since merger. Hence, single-speed ejecta more energetic than $E\approx 10^{50}\,$erg must be slower than $\beta_0\approx 0.5$. Else, radio emission from such ejecta would have peaked before 4.5\,yrs in the radio, at a flux level above $3.3\,\mu$Jy.

\begin{figure*}
    \centering
    \includegraphics[width=\textwidth]{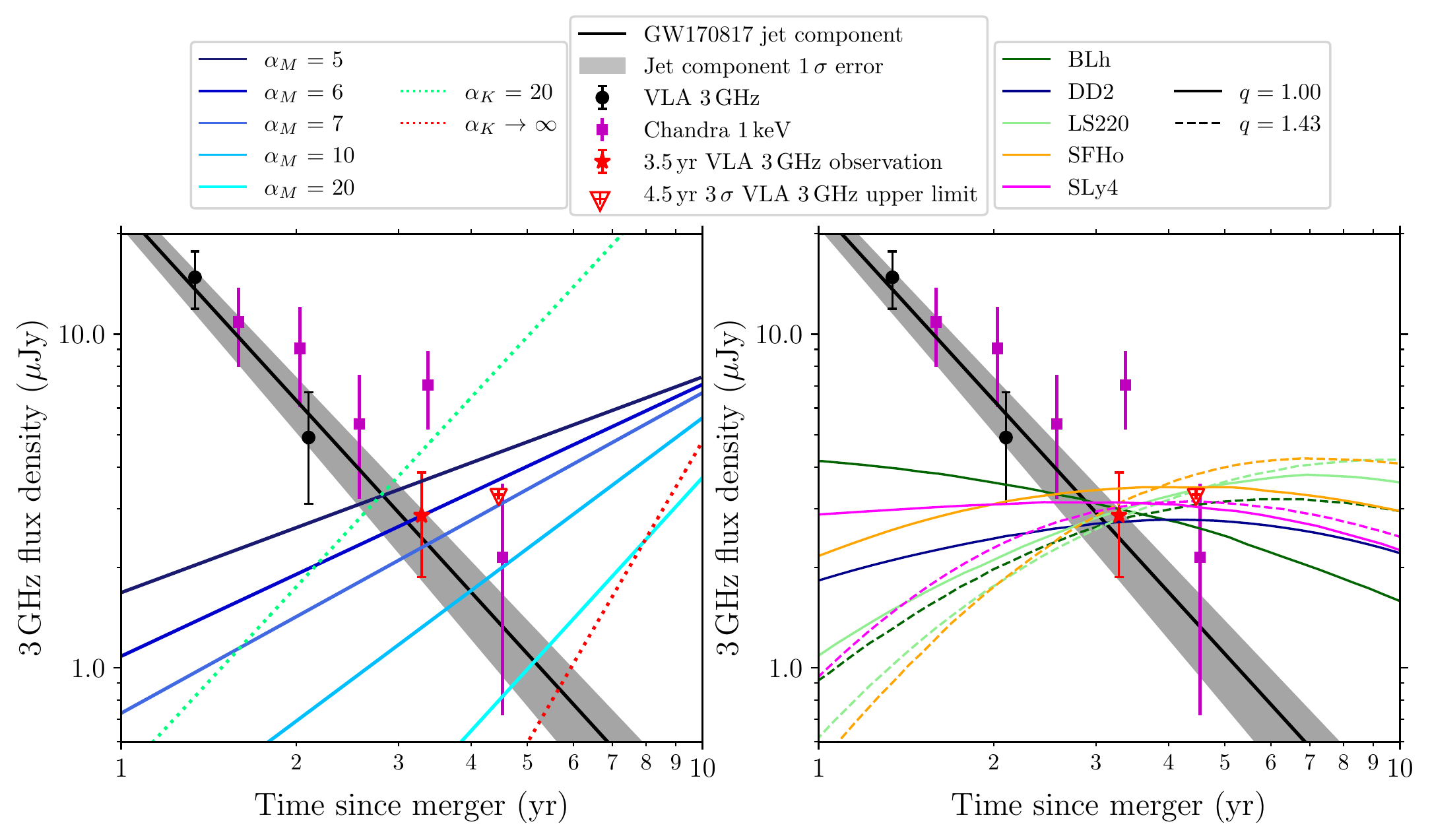}
    \caption{Late-time observations of the GW170817 field at 3\,GHz from  \cite{2021ApJ...922..154M} (black dots),  \cite{2021ApJ...914L..20B} (red star), and from this work (red downward pointing triangle for our $3\sigma$ upper-limit). \textit{Chandra} 1\,keV observations scaled to 3\,GHz with a power-law index of $\beta=-0.584$  \citep[including the latest measurement by][]{Oconnor_GCN} are shown as purple squares \citep[see e.g.][and references therein]{2017ApJ...848L..25H,2017ApJ...848L..20M,2017Natur.551...71T,2022ApJ...927L..17H,2022MNRAS.510.1902T}. We  extrapolate all X-ray data to 3\,GHz  using  $\beta=-0.584$ because there is only marginal evidence for a potential spectral flattening around 3.5 yrs since merger \citep{2021ApJ...914L..20B,2022ApJ...927L..17H} which, however, has not been confirmed in later observations by \citet{Oconnor_GCN}. We compare these observations with predicted kilonova afterglow light curves. \emph{Left:} The solid lines in various shades of blue are the predicted kilonova afterglow light curves as a function of $\alpha$ (see $\S$\ref{eq:kn_rise}) with the assumption that the minimum speed of the ejecta is $\beta_{0}=0.3$, for the parameters $E=10^{51}\,$erg, $p=2.1$, $\epsilon_{e}=7.8\times10^{-3}$, $\epsilon_{\rm B}=9.9\times10^{-4}$, $n=9.8\times10^{-3}{\rm cm}^{-3}$, $d=40\,$Mpc \citep[as in][]{2021ApJ...922..154M} and $\alpha_{M}=5, 6, 7, 10, 20$ (subscript M indicates parameters from \cite{2021ApJ...922..154M}). For comparison, the green and red dashed line show the case $E=10^{51}\,$erg, $\epsilon_{e}=10^{-1}$, $\epsilon_{\rm B}=10^{-3}$, $n=10^{-2}{\rm cm}^{-3}$, $p=2.2$, $d=40\,$Mpc \citep[as in][]{2019MNRAS.487.3914K}, with  $\alpha_{K}=20, \infty$ (subscript K indicates parameters from \cite{2019MNRAS.487.3914K}; see $\S$\ref{sec:disc}); \emph{Right:} Predicted radio light curves of BNS ejecta for different EoS and mass ratios reproduced from \citet{Nedora2021} (see their Figure 4 and Table 2). See $\S$\ref{sec:disc} for discussion.}
    \label{fig:GW170817_lc_with_KN_pred}
\end{figure*}

 Next, in Figure \ref{fig:GW170817_lc_with_KN_pred} we consider the case of a stratified ejecta with an energy-speed distribution described by the parameter $\alpha < \infty$. In this case, we can use our observations to constrain the values of $\alpha$ under specific assumptions on the energy and minimum speed of the ejecta, and of the density and micro-physical parameters. The blue and green curves in the left panel of Figure \ref{fig:GW170817_lc_with_KN_pred} show the rising portion of the predicted kilonova afterglow. Specifically, the shades of solid blue curves assume the parameters $E=10^{51}$\,erg, $\beta_0=0.3$, $p=2.1$, $\epsilon_{e}=7.8\times10^{-3}$, $\epsilon_{\rm B}=9.9\times10^{-4}$, $n=9.8\times10^{-3}{\rm cm}^{-3}$, $d=40\,$Mpc \citep[as in][]{2021ApJ...922..154M}; the dotted green and red curves assume the parameters $E=10^{51}\,$erg, $\beta_0=0.3$, $p=2.2$, $\epsilon_{e}=10^{-1}$, $\epsilon_{\rm B}=10^{-3}$, $n=10^{-2}{\rm cm}^{-3}$, $d=40\,$Mpc \citep[as in][]{2019MNRAS.487.3914K}. 
The radio observations presented here (red downward pointing triangle for our $3\sigma$ upper-limit) constrain $\alpha$ to $\alpha\gtrsim6$ if we assume the parameters as in  \citet{2021ApJ...922..154M}. This is compatible with the constraints one can derive from the X-ray observations reported by \citet{Oconnor_GCN} within the large error bars that affect this X-ray measurement (see the last purple square in Figure \ref{fig:GW170817_lc_with_KN_pred}). For the more general choice of micro-physical parameters \citep{2019MNRAS.487.3914K}, our latest upper-limit is compatible only with the more extreme cases of very steep values of $\alpha$ or with a kilonova blast wave comprised of a single velocity component ($\alpha = \infty$). 

The results presented here can also improve on the constraints discussed in \cite{Nedora2021} regarding the NS EoS. In the right panel of Figure \ref{fig:GW170817_lc_with_KN_pred} we show a plot of the EoS-dependent model radio light curves from \citet{Nedora2021}, compared with the radio upper-limit derived in this analysis. As evident from this Figure, our radio observations at 4.5\,yrs since merger add new constraints on the possible EoSs, disfavoring the softer EoS SFHo (with $p=2.05$, $\epsilon_e=0.1$, $\epsilon_B=0.01-0.001$ and $n=(4-5)\times10^{-3}$\,cm$^{-3}$), as well as the stiffer LS220 (with $p=2.05$, $\epsilon_e=0.1$, $\epsilon_B=0.01-0.001$ and $n=5\times10^{-3}$\,cm$^{-3}$) in moderate mass ratio scenarios ($q\lesssim 1.43$). The SFHo and LS220 EoSs predict the same maximum mass of the cold non-rotating NS, but LS220 correlates with a steeper ejecta energy-speed distribution for $q=1$ \citep{2018ApJ...869..130R}. On the other hand, scenarios like a DD2 EoS with $q=1$, that predict a larger value of the cold, non-rotating maximum NS mass, are still possible.

\section{Summary and conclusion}\label{sec:conc}
In this work, we have presented deep, 3\,GHz observations of GW170817 at $\approx 4.5$\,years since merger. We co-added all the data collected with the VLA via our programs to obtain a deep image of the field, and find no evidence for a re-brightening that can be associated with the kilonova ejecta afterglow in the radio. This confirms our previous results \citep{2021ApJ...914L..20B}.  Overall, the upper-limit we set here and the latest X-ray observations reported by \citet{Oconnor_GCN} reinforce the conclusion that there is no  clear evidence for a late-time re-brightening of the GW170817 non-thermal afterglow emission. Qualitatively speaking, models that envision the emergence of a new emission component at late times, with constant or declining X-ray emission beyond the epoch of $\approx 3.5$\,yrs since merger and without accompanying bright radio emission, could likely still be fit to the data \citep{2022ApJ...927L..17H} but would be very hard to test via further radio (or X-ray) observations of the GW170817 field. 

Kilonova ejecta afterglow models could still be constrained with further radio observations of sensitivity similar to the one reached in this work (and thus not without a substantial investment of observing time). Indeed, the observations presented here tighten the constraints  on the power-law index of the energy-speed distribution of the kilonova ejecta to  $\alpha\gtrsim6$, somewhat steeper than the $\alpha\gtrsim5$ constraint that we obtained at 3.5\,years post-merger \citep[see][]{2021ApJ...914L..20B}, and favor EoS predicting larger values of the cold, non-rotating NS mass for $q=1$ scenarios.

In the future, a radio non-detection at $7.5$\,yrs since merger would constrain the ejecta energy-speed distribution to $\alpha \gtrsim 10 $ for reasonable assumptions on the ejecta parameters. A detection would also facilitate more precise constraints on the possible EoSs (see e.g. BLh in the right panel of Figure \ref{fig:GW170817_lc_with_KN_pred}), though previous considerations regarding  the challenges of pinpointing a specific EoS remain true \citep[see e.g.][]{Nedora2021,2021ApJ...914L..20B}.

\acknowledgements
A.B. and A.C. acknowledge support from NSF AST-1907975. K.P.M. and G.H. acknowledge support from the National Science Foundation Grant AST-1911199. D.L. acknowledges support from NSF grant AST-1907955. The National Radio Astronomy Observatory is a facility of the National Science Foundation operated under cooperative agreement by Associated Universities, Inc. 
\bibliography{references}
\end{document}